# DALSA: Domain Adaptation for Supervised Learning from Sparsely Annotated MR Images

Michael Goetz, Christian Weber, Franciszek Binczyk, Joanna Polanska, Rafal Tarnawski,
Barbara Bobek-Billewicz, Ullrich Koethe, Jens Kleesiek, Bram Stieltjes, Klaus H. Maier-Hein

We propose a new method that employs transfer learning techniques to effectively correct sampling selection errors introduced by sparse annotations during supervised learning for automated tumor segmentation. The practicality of current learning-based automated tissue classification approaches is severely impeded by their dependency on manually segmented training databases that need to be recreated for each scenario of application, site, or acquisition setup. The comprehensive annotation of reference datasets can be highly labor-intensive, complex, and error-prone. The proposed method derives high-quality classifiers for the different tissue classes from sparse and unambiguous annotations and employs domain adaptation techniques for effectively correcting sampling selection errors introduced by the sparse sampling. The new approach is validated on labeled, multi-modal MR images of 19 patients with malignant gliomas and by comparative analysis on the BraTS 2013 challenge data sets. Compared to training on fully labeled data, we reduced the time for labeling and training by a factor greater than 70 and 180 respectively without sacrificing accuracy. This dramatically eases the establishment and constant extension of large annotated databases in various scenarios and imaging setups and thus represents an important step towards practical applicability of learning-based approaches in tissue classification.

*Index Terms*—Brain Tumor Segmentation, Domain Adaptation, Transfer Learning, Automatic multi-modal Segmentation, Random Forest, Glioma

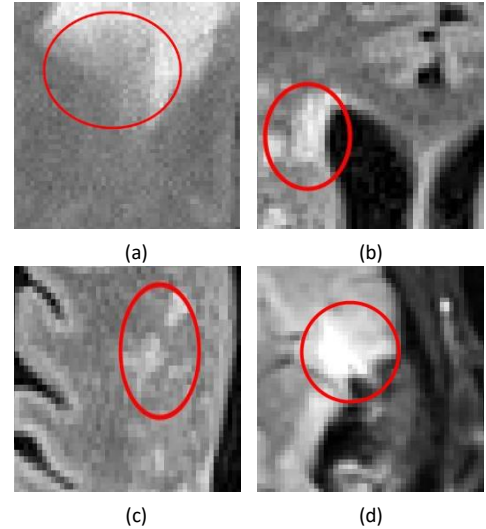

Fig. 1. Parts of T2-Flair images where it is difficult to differentiate between healthy and tumorous tissue. (a) diffuse border between edema and healthy (b) gliosis (c) chronic stroke (d) blood and inflammation

## I. Introduction

Automated approaches for the segmentation of brain tumors allow a time-efficient and objective evaluation of large amounts of data [1]. Current techniques in fullyautomated malignant glioma segmentation use multi-modal data and advanced pattern recognition methods – such as support vector machines (SVM) [2]–[5], neuronal networks (NN) [6], logistic regression [6], or random forests [1], [7] – to learn the appearance of tumorous tissue from manually labeled training data. One important limitation of learningbased approaches is the dependency on a larger amount of complete tumor segmentations. Several factors make it extremely challenging and labor intensive to manually segment malignant gliomas and to create the required training data [8]–[10]. Gliomas invade the surrounding tissue rather than displacing it. This leads to blurry and unclear borders (Fig. 1a). This is aggravated by the fact that other tissue types, such as gliosis (Fig. 1b), stroke (Fig. 1c), inflammation, and the presence of blood (Fig. 1d) can all have a similar appearance in the MR images. It is therefore difficult to create a complete segmentation of tumorous and non-tumorous tissue without leaving a substantial amount of voxels labeled as *unknown*. The size of these tumors, their irregular shape, and the heterogeneous growth patterns add to the difficulties in this complex and error-prone task. Consequently, Mazzara *et al.* report an intra- and inter-observer variability of 20% and 28% respectively for gross tumor segmentation [11]. Similar Menze *et al.* reported interrater-variances of $85\pm8\%$ for whole tumor segmentation and $74 \pm 13\%$ for active tumor segmentation [12]. This implies that the commonly used training data and reference segmentations are not flawless.

Some previous studies completely avoided supervised learning on basis of manual segmentations for these reasons, for example by analyzing brain symmetry [13] or by incorporating atlas information in combination with geometric and spatial priors [14]. Parisot *et al.* employed model-based

M. Goetz, C. Weber, J. Kleesiek and K.H. Maier-Hein are with the Junior Group Medical Image Computing, German Cancer Research Center, Heidelberg, Germany.
F. Binczyk and J. Polanska are with the Data Mining Group, Institute of Automatic Control, Silesian University of Technology, Gliwice, Poland.
R. Tarnawski and B. Bobek-Billewicz are with the Maria Sklodowska-Curie Memorial Cnacer Center and Institute of Oncology Gliwice, Poland.
U. Koethe and J. Kleesiek are with the Heidelberg Collaboratory for Image Processing (HCI), University of Heidelberg, Germany
B. Stieltjes is with the Department of Radiology, University Hospital Basel, Switzerland

analysis in a complementary way by combining it with a learning-based tumor annotation [15].

To increase the quality of manual segmentations that can be used for training and evaluation of supervised approaches, Pedoia et al. proposed a tool for supporting labeling of glial tumors [13]. They showed that the use of their tool reduces labeling errors and results in more consistent segmentations. Another way of improving segmentation quality is to fuse multiple segmentations. Warfield et al. showed that a fusion based on 'Simultaneous truth and performance level estimation' (STAPLE) increases segmentation quality and yields more consistent segmentations [16]. While these approaches can reduce the variance in the segmentations, they cannot solve the problem of ambiguous image information. The experts are still confronted with a significant uncertainty about the true tissue borders. Also, the problem remains that the annotations are highly labor-intensive: it takes up to an hour to create a high-quality segmentation [11], [12].

MRI does not have a standardized measurement value per voxel, i.e. it is not quantitative, as it is for example the case for computed tomography (CT). The image intensity is given in arbitrary units and cannot be easily calibrated without a ground truth measure. Thus, both gray scale images and MRI-derived semi-quantitative values (like apparent diffusion coefficient ADC, or fractional anisotropy FA) exhibit a substantial variability; a variation of 15% is not extraordinary (e.g. [17] for the ADC). Potential sources of variation can include varying internal value ranges, slightly varying sequence implementations, or varying noise characteristics between scanners or different manufacturers. In practice, the annotations needed for classifier training must therefore be repeated when the scanner hardware or MR sequence settings change. Also, the introduction of new MR sequences or a reconfiguration of imaging protocols may require a novel training set to be reestablished. Since precise manual labeling is time-consuming and difficult, this is a very expensive task, which reduces the likelihood of these methods being incorporated in a clinical setting.

For this reason, the use of incomplete segmentations was introduced in previous work to avoid long labeling times in generating the training data. Verma et al. used a nearly complete segmentation that omitted ambiguous areas [5]. With these incomplete segmentations they performed an intrapatient segmentation with Gaussian mixture models (GMM) and trained an SVM classifier for inter-patient segmentation. Kaus et al. segmented low-grade brain tumors by combining brain atlases and classifiers [18]. A first voxel-based segmentation was registered to an atlas and this data was then used to refine the segmentation. This is repeated iteratively and the classifier is re-learned every time. The whole method was initialized with few manually selected and not necessarily connected voxels. Although these approaches used incomplete segmentations to train classifiers for tumor detection, none of them investigated or corrected for the influence of the incomplete segmentations.

We propose a new approach that allows learning from partial, incomplete annotations, which sparsely represent each label class and ensure unambiguous assignments of labels to voxels. The approach yields complete annotations of MR images but keeps the necessary training annotations sparse, unambiguous and thus fast to perform compared to full annotations. The resulting sampling selection error introduced by the sparse and unambiguous annotation leads to differences between the training and test data distributions, i.e. $p_{\text{train}}(x) \neq p_{\text{test}}(x)$. As suggested by Cortes et al. we consider the non-i.i.d sampled training data as the training domain and the full images as the testing domain [19], thus treating the problem as a special instance of a domain adaptation (DA) problem. This allows us to correct for the sampling bias using DA techniques which optimize performance on one domain, given training data that is from a different domain [20]. Our training and test domains differ in the sampling and not in their origin as it is for example the case in the work of Heimann et al., who trained classifiers for ultrasound transducer localization from synthetic images [21]. We show that our method returns similar results to traditionally trained classifiers on a set of clinically acquired images, even though the traditional classifiers were trained directly on the gold standard reference segmentations. Compared to our previous work [22], this paper offers a more detailed analysis and experimental protocol.

With sparse annotations becoming feasible as means to train automated classifiers, the generation and continuous expansion of large annotated data sets for various kinds of lesions and imaging setups comes within reach, allowing a workflow that is fully integrated into a clinical setup. The fast labeling strategy also comes with the added value of allowing the differentiation of multiple tissue classes. While this paper focuses on the segmentation of malignant gliomas in MR images, the presented method can also be applied to other applications. It might be used to improve other learning-based approaches in any scenario involving segmentation or imagebased tissue characterization.

II. METHODS

To reduce the labeling time necessary for creating training data for automatic tumor segmentation, we propose the annotation of sparse and unambiguous regions (SUR) instead of segmenting the complete images. Unlike learning from complete annotations (LCA), learning from sparse annotations (LSA) introduces a sampling bias. We propose to correct this error with domain adaptation, which we refer to as domain adaptation for Learning from Sparse Annotations (DALSA). The different methods used for annotating, sampling, and using training data are summarized in Fig. 2. In this work, the sparse annotations (SURs) were typically located in one or two slices of the image and covered approximately 1% of the brain voxels. Fig. 3 gives some examples of SURs and Table I shows different annotation strategies that were applied to create SURs.

### A. Domain Adaptation

A basic assumption in machine learning is that the training data are independent and identically distributed (i.i.d.) [23]–[25]. However, if only small areas of an image are used for training, this assumption becomes invalid and a sampling selection bias occurs [26]. The distribution of features $x$ and labels $y$ will be different in the observations $(x,y)$ processed during training and testing, i.e. $P_{\text{Train}}(x,y) \neq P_{\text{Test}}(x,y)$. This will lead to classifiers with non-optimal decision boundaries, since some features may be overrepresented while others are underrepresented.

Fig. 4 shows a simplified example to demonstrate the effect of sampling selection error and domain adaptation. The probability for a combination of feature vector and label $P(x,y)$ is affected by a sampling bias. This probability can be written as:

$$P(x,y) = P(y \mid x) \cdot P(x). \qquad (1)$$

A theoretical assumption often made in domain adaptation is that the meaning of a feature is the same in the training and test domains, e.g.

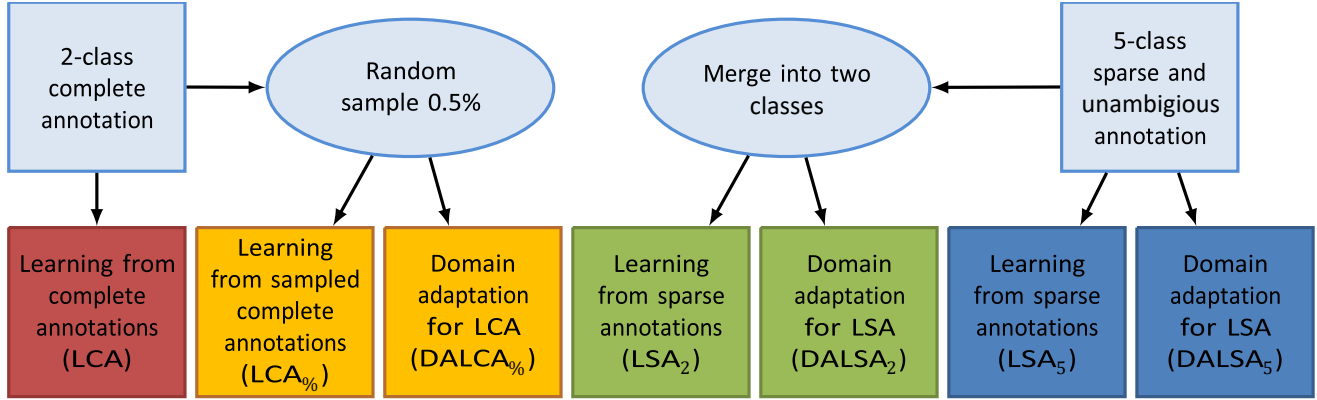

Fig. 2. Different methods used for annotating, sampling, and using training data for supervised learning. Most state-of-the-art approaches make use of LCA or LCA$_\%$, which require a complete annotation of the data and differ in their sampling strategy. We propose the use of sparsely annotated training data (LSA and DALSA) to reduce the annotation time. The sparse annotations of 5 tissue classes were either treated separately or merged to two classes ('healthy' and 'fluid' were merged to 'healthy', 'edema', 'active' and 'necrosis' were merged to 'tumorous')

$$P_{\text{Train}}(y \mid x) = P_{\text{Test}}(y \mid x). \qquad (2)$$

TABLE I
LABELING STRATEGIES

| Type | Description | Diameter | Location |
|---|---|---|---|
| Main | 1-3 SUR per class | rater dependent | covering bordering as well as central tissue areas |
| Type 1 | 1 SUR per class | 6–14mm | arbitrarily varying |
| Type 2 | 3 SURs per class (different slices) | 6–14mm | covering bordering as well as central tissue areas |
| Type 3 | 3 SURs per class (different slices) | 6–14mm | covering central tissue areas only |
| Type 4 | 3 SURs per class (different slices) | 6–14mm | covering bordering tissue areas only |

Description of different SUR labeling strategies. A complete set of SURs was created for each strategy.

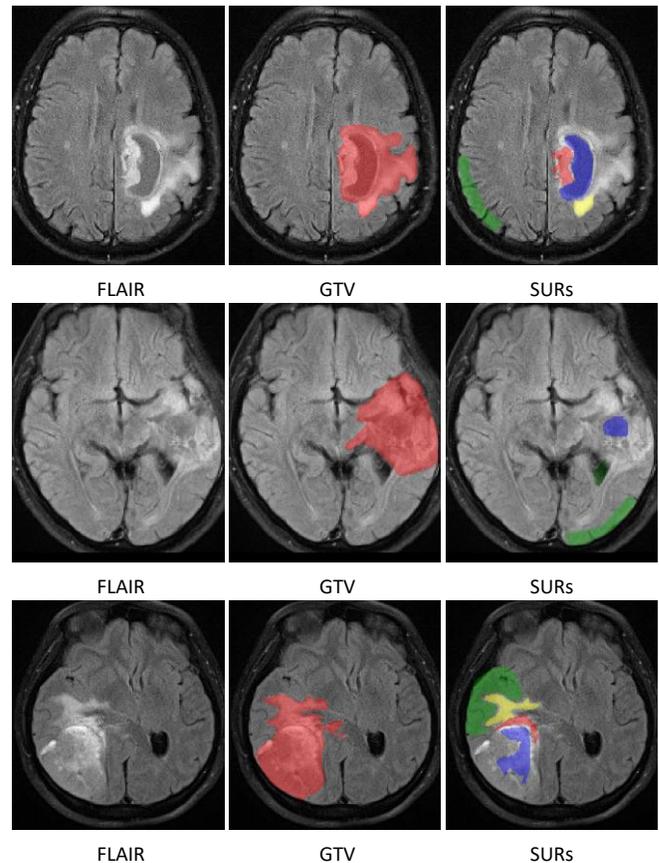

Fig. 3. Exemplary axial slice of the annotations for 3 patients. Left is the FLAIR without annotation, in the middle the complete manual annotation and on the right the SUR annotation. The color coding for the SUR is green: 'healthy', yellow: 'edema', red: 'active tumor', blue: 'necrosis'. Labels not shown in the images are annotated in a different slice.

Huang *et al.* [27] showed that techniques that depend on this assumption are useful even if this assumption is only partially fulfilled. The remaining difference between the distributions in the training and test data is the probability of a given feature

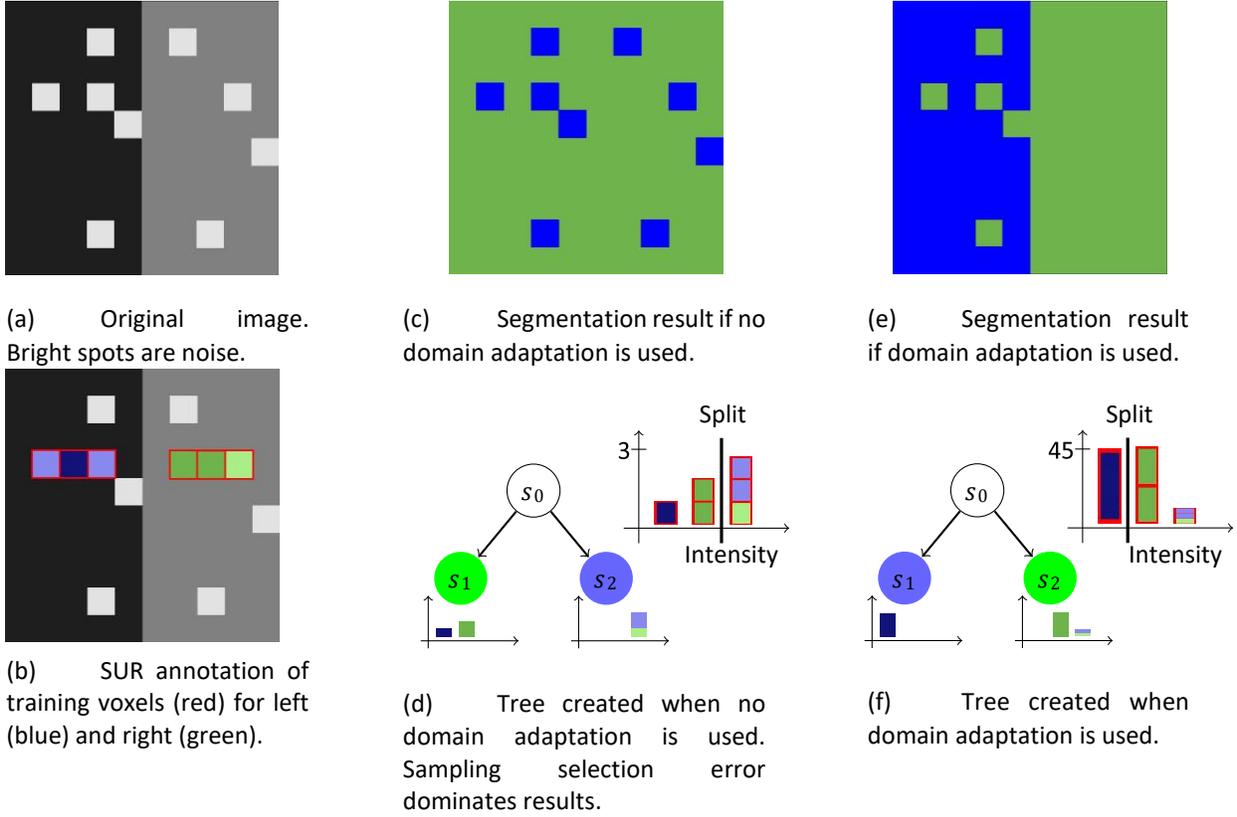

(a) Original image. Bright spots are noise.

(b) SUR annotation of training voxels (red) for left (blue) and right (green).

(c) Segmentation result if no domain adaptation is used.

(d) Tree created when no domain adaptation is used. Sampling selection error dominates results.

(e) Segmentation result if domain adaptation is used.

(f) Tree created when domain adaptation is used.

Fig. 4. Simplified example to demonstrate the effect of sampling selection error and domain adaptation. (a) The given image with 100 pixels is classified into left and right using intensity as feature. On both sides salt noise (bright pixels) simulates noise in the multidimensional data. (b) For training data, SURs are used. A single tree with a tree depth $T = 1$ is used as classifier. (c) shows the segmentation result with the original image if no domain adaptation is used; (d) shows the resulting tree that has a false split due to the noise pixels. (e) gives the segmentation result if domain adaptation is used, (f) gives the tree from the corrected data. The number of pixels at each node differs from the number of the pixels within the SURs because the classifier uses the number of pixels multiplied by a weight. For example, there are nine bright pixels in the given image and the SURs cover three of them. Therefore the weight for bright features is $w(\text{BRIGHT}) = 9 \div 3 = 3$.

vector

$$P_{\text{Train}}(x) \neq P_{\text{Test}}(x). \qquad (3)$$

Shimodaira [28] calls this situation covariate shift. He suggests compensating the difference by weighting each sample with the density ratio of the feature vectors during training:

$$w(x) = \left(\frac{P_{\text{Test}}(x)}{P_{\text{Train}}(x)}\right)^{\lambda}. \qquad (4)$$

The ratio $w(x)$ is high for observations occurring often within test data and seldom within training data, while $w(x)$ is low for observations that are rare in the test data but frequent in the training data. In our case this means that labeled voxels (training data) that are typical for the entire image (test data) receive more emphasis than untypical voxels. The relaxation coefficient $\lambda \in [0..1]$ was introduced by Shimodaira to control the effect of the weights [28]. If $\lambda = 0$, the weights do not have any effect and for $\lambda = 1$ the effect of the weights is maximized. The best value for $\lambda$ depends on the number of available training points and the used classifier; in general $\lambda$ needs to be smaller for small training sets. We set $\lambda$ to 1 because, being voxel-based, our training base is rather large. This choice was further evaluated with an experiment.

Since $w(x)$ is unknown for most applications, it is usually estimated. There are several ways to do this and [29] gives an overview for the most common methods. We chose the approach of assessing $w(x)$ by estimating the probability of whether an observation with feature vector $x$ belongs to the training or the test data [30]. If observations that are used for the training data are labeled $z = 1$ and observations that are tested $z = 0$ then $w(x)$ can be estimated by

$$\begin{aligned} \hat{w}(x) &= \left(c \cdot \frac{\hat{p}(z=0 \mid x)}{\hat{p}(z=1 \mid x)}\right)^{\lambda} \\ &= \left(c \cdot \frac{1 - \hat{p}(z=1 \mid x)}{\hat{p}(z=1 \mid x)}\right)^{\lambda}. \end{aligned} \qquad (5)$$

We estimate the probability $\hat{p}(z = 1 \mid x)$ by training a logistic regression classifier (LRC). For this purpose, each voxel within a SUR is labeled as training data, i.e. $z = 1$. Additionally, all voxels that belong to the brain are labeled as test data, i.e. $z = 0$. Thus the voxels that belong to the SURs appear twice: once within the training data and once within the test data. These data are then used to train the parameter function of the logistic regression $\theta(x)$, which can then be used to estimate the required probability by:

$$\hat{p}(z=1 \mid x) = \frac{1}{1+\exp(-\theta(x))} \quad (6)$$

The estimation of $w(x)$ can be further simplified by incorporating equation 6 in equation 5:

$$\hat{w}(x) = (c \cdot \exp(-\theta(x)))^\lambda. \quad (7)$$

This allows a fast calculation of the weights and eliminates the need for division, which increases the numerical stability. We used generalized linear models with a logit function as link function and a binominal distribution to fit the logistic regressor [31]. The advantage of this method is the parameterfree training of the logistic regressor, which makes the whole algorithm more robust and easier to use.

The weights are calculated patient-wise, e.g. for each patient the SURs are created and then the weights for this patient are calculated. Therefore the weights for a patient are independent of other patients and new patients can be added to the training data without recalculating the weights within the existing training data. Also, no full tumor segmentation is necessary.

The constant $c$ can be used to control the influence of each image volume during the training without changing the relations of voxels which belong to the same image. The sum of weights of all SUR voxels is $c \cdot n_{\text{Test}}$ ($n_{\text{Test}}$ is the total number of voxels in the brain mask, see appendix A for a mathematical derivation). A common approach is to set $c = \frac{n_{\text{Train}}}{n_{\text{Test}}}$, with $n_{\text{Train}}$ being the number of voxels in the SURs [29]. This normalizes the sum of all weights to the number of training points. However, in our case this would mean that the contribution of an image to the overall training depends on the size of the SURs, i.e. an image with large SURs would have more impact on the final classifier than an image with small SURs, although the latter might contain more valuable information. We therefore set $c = 1$. In this case, the impact of an image is determined by $n_{\text{Test}}$, as it would be in a standard classifier training scenario.

### B. Observation-weighted Classifier

The classification in our main experiments is based on random forests [32], [33]. Random forest-based methods have previously achieved promising results in brain tumor segmentation [1], [7]. Since the original random forest definition does not support observation weights, we used a variant that is similar to the random forest implementation in the python scikit module [34]. Here, the prediction algorithm itself is not modified. An unseen observation is passed down the decision trees based on binary tests within each node of the tree until it reaches a leaf node. The prediction is then obtained by majority voting of all trees. The training of observationweighted random forests is also similar to the original version. At each node, the best split within a random set of features is determined based on an impurity measurement. All data are then split into two groups, which are used to train the child nodes. This is repeated until the maximum tree depth is reached or only one label type is left.

The major difference to the canonical random forest is how the impurity is calculated. We used the Gini Impurity $I$ [35]:

$$I(V) := 1 - \sum_{y \in Y} P_V(y)^2 \quad (8)$$

The class probability $P_V(y)$ is usually calculated from the number of observations with this label divided by the overall number of observations. If observation weights are used, $P_V(y)$ is calculated using the sum of the weights instead of the number of observations.

The forests in our experiments consisted of 1000 trees; the number of features at each node was set to the square-root of the number of all features (i.e. 4). The minimum sample size at each leaf node was set to 1, the noise reduction being instead achieved by limiting the maximum tree depth. To account for the different levels of noise and amount of data, we ran multiple runs of our experiments with different tree depths and used the optimal tree depth for each approach.

While we used random forests in our main experiments, our method can be used with any classification algorithm that allows for the incorporation of observation weights. We demonstrated this in an additional set of experiments on basis of weighted SVMs as described by Yang *et al.* [36]. We chose a non-linear radial basis function kernel (Gaussian) and the Karush-Kuhn-Tucker stop criterion [37]. The noise sensitivity is regularized using the cost parameter $c$.

## III. EXPERIMENTS

### A. Dataset I

Dataset I consists of a longitudinal study including 19 patients with high-grade glioma. For each patient, a T1-weighted image with contrast agent (T1C), T2-weighted FLAIR image (FLAIR), and a diffusion tensor image (DTI) reconstructed from a diffusion-weighted image (DWI) were available. All images were acquired during clinical routine on a single 1.5T Siemens Avanto (Siemens Health Care, Erlangen, Germany) using a standard protocol with a duration of less than 20minutes per examination. T1C and FLAIR images were acquired with an in-plane resolution of $0.55 \times 0.55$mm and $0.65 \times 0.65$mm, respectively, and 6mm between slices. DWI parameters were: single-shot spin echo EPI sequence, echo time = 95ms, repetition time = 3.6s, slice thickness of 4mm, and pixel spacing of $1.8 \times 1.8 \times 5.2$mm. Two shells (b = $1000, 2000$s/mm$^2$) with 48 directions were acquired. Several commonly used parameter maps were calculated from the DTI: fractional anisotropy (FA), relative anisotropy (RA), axial diffusivity (AD), radial diffusivity (RD), clustering anisotropy (CA), and mean diffusivity (MD) [38]. All parameter images were calculated with and without free water elimination (FWE) [39]. A free-water map (FW) and a b0-image (B0) were also included. Fig. 5 shows examples of the image contrasts.

A single point in time (13 pre-operative and 6 postoperative) was selected for each patient for the experiments. We selected either the last point in time before or the last available point in time after an operation.

All images were rigidly registered intra-patient-wise to the FLAIR image and then resampled to a common resolution of 1mm × 1mm in-plane. The slice thickness was set to 3mm, compromising between resampling artefacts and the number of slices that needs to be labeled. A semi-automatic brain-mask was created and B0, T1C and FLAIR were normalized to the most frequent gray value inside the brain mask and a standard deviation of 1.

An expert radiologist segmented the GTV manually based on T1C and FLAIR. We defined the gross tumor volume as the area that covers edema, contrast-enhancing areas and necrosis. To reduce segmentation errors, multiple refinement runs were carried out. We used Tumor Progression Maps (TPM) [40] to ensure the consistency of segmentations with the other time steps of the same patient. TPMs highlight areas in which the tumor outlines change over time. Cases where such changes were caused by inconsistencies in the segmentation could thus be quickly identified and corrected. The time required for segmenting a single time point for a patient was more than six hours. Although significantly longer than clinically applied segmentation procedures [11], this time was taken since the resulting segmentations were used for training and validation of our classifiers.

In addition, three raters (one expert radiologist and two medical students) segmented independent sets of SURs (c.f. Table I, 'Main'). Each rater was blindfolded to the complete tumor segmentation and the SURs created by the other raters. SURs were defined for each of five different clinically relevant tissue classes including high proliferative tumor parts (active tumor, e.g. as potential target for biopsies), necrosis (e.g. as an indicator of tumor grade and poor target for biopsies), and low proliferative tumor parts (edema, e.g. as part of the peripheral tumor border) in addition to healthy tissue and cerebrospinal fluid (CSF). The task of the raters was to annotate small areas which are typical for each tissue class. If possible, areas close to tissue borders should be included if they were clearly distinguishable from the neighboring class. No other restriction in terms of size, number of ROIs per patient, relative location, or number of annotated slices was made. It took less than five minutes to create these small 2DROIs, which were usually located in one or two single slices of an image. Fig. 5 shows an example of a complete gross tumor segmentation and the SURs. The mean coverage ratio of segmented voxels to brain voxels for the SURs created by rater 1, 2 and 3 were 0.53% ± 0.23%, 0.41% ± 0.11%, and 0.18% ± 0.05%, respectively. The minimum and maximum coverage ratios were 0.24%, 0.17%, and 0.08% and 1.22%, 1.16%, and 0.33%, respectively. On average, 2.6% ± 1.5%, 1.6% ± 1.1%, and 1.0% ± 0.5% of the tumorous tissue were covered by SURs.

To analyze the effect of varying SUR placement strategies, a medically trained expert created four additional different sets of SURs using different labeling strategies (c.f. Table I, 'Type 1'-'Type 4'). The mean coverage ratio of segmented voxels to brain voxels for these SUR sets was 0.2%±0.1% (maximum: 0.63%, minimum: 0.06%). The SURs of Type 1 were the smallest (0.12% ± 0.03%) and those of Type 2 the largest (0.28%±0.10%). Those of Type 3 and 4 were similar in size scattering around 0.18%. The mean tumor coverage ratio was 1.3% ± 0.1% across all types.

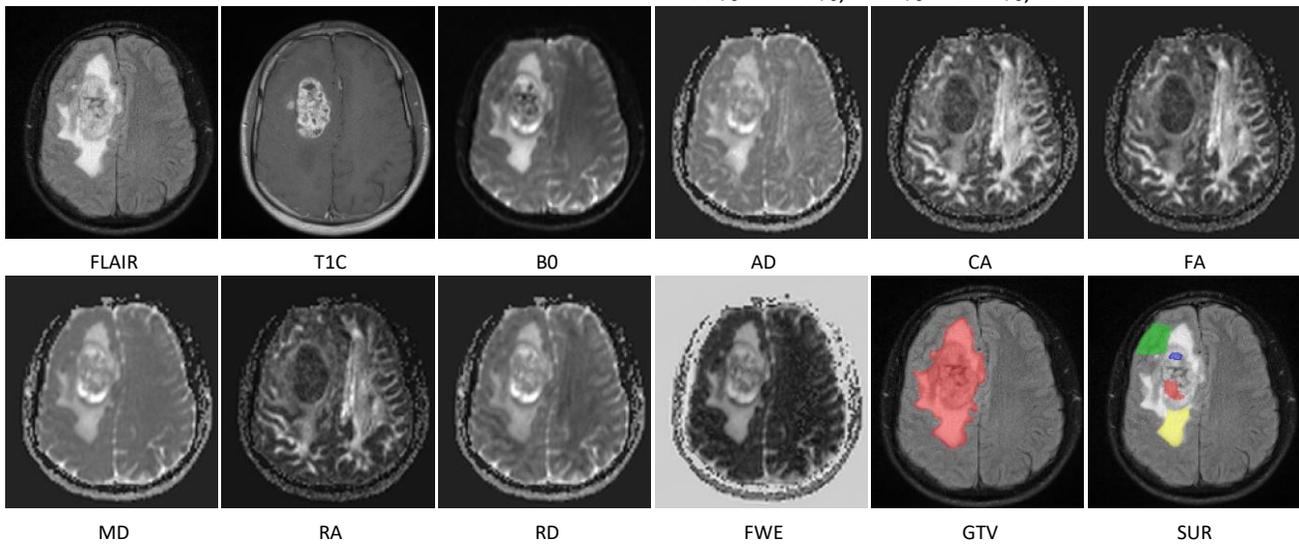

Fig. 5. Exemplary axial slice of the available contrasts except the free-water-corrected contrasts and the available complete and sparse annotations. The color coding for the SUR is green: 'healthy', yellow: 'edema', red: 'active tumor', blue: 'necrosis'. 'Fluid' is not segmented in the shown slice.

*B. Dataset II*

The BraTS 2013 challenge dataset was used as second dataset to compare our approach to the state of the art in tumor segmentation. The dataset consists of 20 training and 10 test subjects. For every subject a T1-weighted (T1), a contrast-enhanced T1-weighted (T1C), a T2-weighted (T2), and a T2-weighted FLAIR (FLAIR) image is given. All images were already registered intra-patient-wise and resampled to a common resolution. Semi-manually created annotations of the training set on the basis of 4 raters are provided. The remaining 10 subjects are used to compare the different algorithms; the segmentations for these patients are not public. The mean inter-rater variance was reported at 85% ± 8% for the class 'whole tumor', i.e. GTV based on our definition [12]. Total annotation time per rater was about one hour, resulting in a total annotation time per segmentation of four hours [12]. Similar to dataset I, we defined additional SUR segmentations for DALSA.

*C. Experimental Protocol*

Classifiers were trained in leave-one-patient-out experiments. The quality of the obtained segmentation on the left-out patient was evaluated on the basis of the manually annotated ground truth using the well-known DICE score [41] as well as the sensitivity (true positive rate) and specificity (true negative rate). The ground truth contained only the labels 'tumorous' and 'healthy'. Thus, in the five-class automatic segmentation, the labels 'healthy' and 'fluid' were relabeled 'healthy', while the labels 'edema', 'active' or 'necrosis' were relabeled 'tumorous' prior to evaluation. In all our experiments – except the generation of Fig. 8 – the decision threshold of the classifiers was left at 50% for the two class problem. The decision threshold was not affected by adding the class weights used by our method. The following three different setups were used in our experiments.

*Setup I* consisted of dataset I in conjunction with weighted random forests. The feature vector $x_i$ for a voxel $v_i$ of this setup consisted of the gray values of all available images, i.e. FLAIR, T1C, B0, AD, CA, FA, MD, RA, RD, AD-FWE, CAFWE, FA-FWE, MD-FWE, RA-FWE, RD-FWE, and FW.

On the basis of setup I, we assessed seven different methods for annotating, sampling, and using training data (cf. Fig. 2). As a reference, three of those methods were trained using the ground truth GTV segmentations as training labels. Two different sampling strategies were applied: sampling of all labeled voxels (Learning from Complete Annotations, LCA) and random sampling of the labeled voxels at $0.5\%$ ratio (similar to SUR coverage ratio of the expert rater). The randomly sampled training data were used with (DALCA$_\%$) and without (LCA$_\%$) domain adaption. The reference methods were compared to classifiers trained on SURs created by the expert radiologist, either using (DALSA) or not using (LSA) domain adaptation. The SURs differentiated five different tissue classes while the ground truth segmentations only differentiated tumorous from healthy tissue. It was therefore necessary to fuse tissue classes to allow direct comparison of the different methods. Indices indicate the number of classes that were used during training (e.g. LSA$_2$: fusion of labels before training; LSA$_5$ fusion of labels after training in predicted images).

The influence of $\lambda$ was evaluated by conducting leave-onepatient-out experiments for DALSA$_2$. We set the maximum tree depth to 4, the minimum sample size at each leaf node to 1, and the maximum number of evaluated features at each node to 4. We then varied $\lambda$ between 0.0 and 1.0. The influence of altering SUR annotations was evaluated in two experiments: First, LSA$_2$ and DALSA$_2$ classifiers were trained on SURs with varying annotation strategies (cf. Table I, 'Type 1'-'Type 4'). Second, we compared the expert's influence on the resulting segmentation quality by training LSA$_2$ and DALSA$_2$ classifiers on SUR sets created by the expert rater and compared these with sets created by two student raters. We also applied majority voting to computed combined results of all raters.

*Setup II* was similar to setup I, but used weighted SVM instead of weighted random forests. On the basis of setup II, we assessed whether SVM-based classification can also profit from DALSA. We conducted leave-one-patient-out runs at varying cost settings between 0.01 and 0.08 and compared the results obtained by LSA$_2$ and DALSA$_2$.

*Setup III* was used to evaluate the performance of DALSA on the basis of the BraTS 2013 challenge data (c.f. dataset II), in contrast to other segmentation approaches that are trained on complete segmentations. For the experiments, we adapted the pipeline of Kleesiek *et al.*, who scored third on the on-site BraTS 2014 challenge [42]. The same preprocessing, features, and post-processing as in the original work were used. Only the sample selection was varied. Instead of randomly drawing a fixed number of samples for each tissue class (which corresponds to our LCA$_\%$ training scheme), we used LSA or DALSA on the basis of the SURs that we had defined.

On the basis of setup III, we assessed whether it is possible to integrate our approach in another existing tumor segmentation pipeline and compared the obtained results for LSA and DALSA with state-of-the-art methods that were trained on complete manual annotations on the basis of the ongoing BraTS 2013 challenge.

*D. Implementation*

We implemented setup I using MITK – a widely used C++ framework for medical imaging [43], [44] – which allows loading, processing and visualization of medical images. We added support for Vigra for random forest capability [45]. Both training and prediction of the random forests were performed multi-threaded on the basis of OpenMP. The complete workflow has been released open-source as part of MITK (www.mitk.org). The SVM-based classification for setup II was implemented in Matlab. For setup III we adapted the implementation by Kleesiek *et al.* [42] that is based on Vigra python bindings. Calculations were partially carried out on infrastructure of GeCONiI (POIG.02.02.01-24-099/13).

## IV. Results

### A. Setup I

Fig. 6 shows the obtained DICE scores and ROC analysis for the different methods assessed. Table II lists the corresponding uncorrected statistical significance values on the basis of the Wilcoxon signed rank test. Fig. 7 provides some exemplary qualitative results. The proposed domain adaptation could effectively reduce the drop in segmentation quality caused by learning from sparsely annotated data. DALSA results did not significantly differ from the results obtained by LCA$_{\%}$, which is a commonly applied sampling strategy in other studies but requires complete annotations. It was possible to further increase the quality of the final segmentation by merging the segmentations obtained from different experts. The merged DALSA results did not significantly differ from LCA ($p$ = 0.084). There was also a significant increase in segmentation quality when applying domain adaptation to LCA$_{\%}$. The extent of the effect, however, was very small (median DICE difference 0.0008, mean $6.23 \times 10^{-5}$).

Fig. 8 demonstrates the effect of domain adaptation on the classification results. The LSA$_2$ DICE scores are plotted over a moving decision threshold (blue curve) and should optimally exhibit a bell-shaped curve with its maximum at 50%. The SUR-based sampling bias, however, lead to a skewed curve with suboptimal classification results. DALSA corrected for this effect and yielded a more bell-shaped curve (Fig. 8a, red curve). Fig. 9a shows the performance of DALSA under different SUR labeling strategies. DALSA significantly outperformed LSA (p ≤ 0.001) in all cases. Similarly, DALSA outperformed LSA regardless of which expert labeled the data (Expert 1: p = 0.015, student 1: p = 0.007, student 2: p = 0.001, c.f. Fig. 9b). DALSA performance was always comparable to LCA$_{\%}$. Fig. 10 shows the influence of the relaxation coefficient $\lambda$.

### B. Setup II

Fig. 9c shows the results obtained by SVM-based classification. Again, DALSA outperformed LSA in all experiments (p ≤ 0.005). DALSA results on the basis of SVM did not significantly differ from DALSA results on the basis of random forests (p-values between 0.08 and 0.28).

### C. Setup III

BraTS 2013 challenge results are shown in Fig. 7g. On the 10 test datasets DALSA yielded a visible increase in segmentation accuracy with respect to DICE score (0.84 to 0.86) and Sensitivity (0.74 to 0.78) for GTV[1]. The Positive Predictive Value was reduced from 0.94 to 0.93. The resulting segmentation quality was similar to those achieved in the original approaches of Kleesiek et al. and Peres et al. that were trained on complete segmentations (reported DICE for both was 0.86). While sensitivity was clearly lower than in both previous approaches (0.91 and 0.87), this was compensated by the Positive Predictive Value (0.83 and 0.85 in the previous approaches). P-values were not calculated due to the small number of test subjects.

### D. Annotation Time and Performance

The mean times required both for creation of the training data and for training and application of the forests are provided in Table III. The SUR-based training was faster than training with sampled or complete data. Since less data needed to be labeled and labeling was more straightforward, the sparse annotation took less than five minutes per patient (for all annotation strategies), while the full annotation took more than six hours (a reduction of labeling time by a factor of more than 70).

## V. Discussion

We presented a new approach that allows training of classifiers in automatic tumor segmentation using easy-to-annotate SURs instead of complete segmentations without sacrificing segmentation accuracy. The proposed domain adaptation technique correctly compensates for sampling selection errors and yields results that are comparable to state-of-the-art methods

TABLE II
STATISTIC SIGNIFICANCE BASED ON WILCOXON SIGNED-RANK TEST

|  | LCA | LCA$_{\%}$ | DALCA$_{\%}$ | LSA$_2$ | DALSA$_2$ | LSA$_5$ |
|---|---|---|---|---|---|---|
| LCA$_{\%}$ | **.001** ↑.007 |  |  | X |  |  |
| DALCA$_{\%}$ | **.001** ↑.007 | **.017** ←.001 |  |  | X |  |
| LSA | **.003** ↑.087 | **.015** ↑.077 | **.012** ↑.078 |  |  | X |
| DALSA$_2$ | **.035** ↑.019 |  | .409 | .387 | **.015** ←.038 | X |
| LSA$_5$ | **.003** ↑.046 |  | **.028** ↑.039 | **.023** ↑.040 | .121 | .191 X |
| DALSA$_5$ | **.017** ↑.019 | .220 | .220 | **.003** ←.051 | .251 | **.003** ←.027 |

Uncorrected p-values of Wilcoxon signed-rank test indicating differences in segmentation results based on the DICE score for the gross tumor volume. $p$ 6

---

[1] LSA and DALSA-results were obtained with our implementation of the approach of Kleesiek *et al.* [42]. They differ from the original results due to the different training setting.

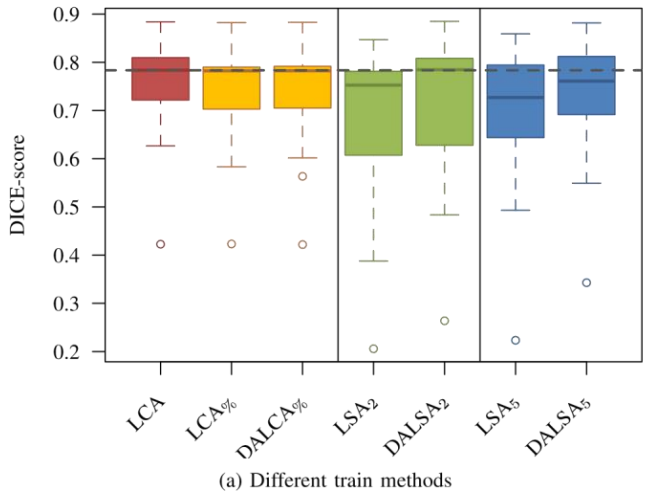

(a) Different train methods

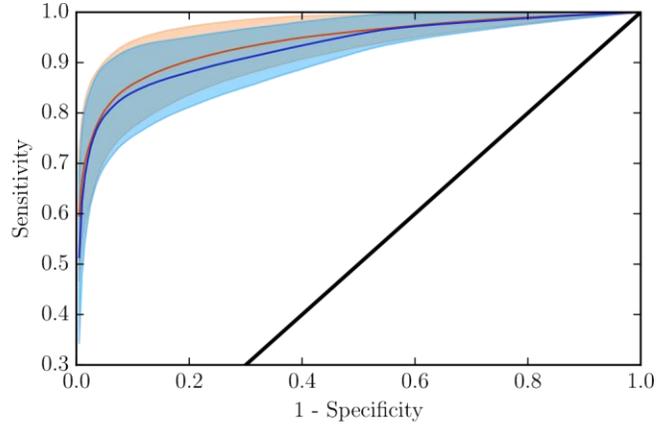

(b) ROC-curve for LSA$_2$ (blue) and DALSA$_2$ (red)

Fig. 6. The results of the leave-one-patient-out experiments using setup I. (a): Boxplots showing the results grouped by classifier scheme. The left section shows the results for classifiers that were trained on complete segmentations. The middle and right sections show results for classifiers trained on SURs using two and five different tissue classes respectively. (b): ROC-curves for LSA$_2$ (blue) and DALSA$_2$ (red). The curves where obtained by varying the decision threshold of the classifier and then calculating the mean (solid) and standard deviation (colored area).

.05 is shown in bold. The absolute difference of the group median DICE scores is shown below the significant p-values. Arrows point to the group with higher median score. For example, LCA performs significantly better than LCA$_\%$.

TABLE III
RUNTIMES

| Method | Training time | Prediction time | Tree depth |
|---|---|---|---|
| LCA | 192.5±1.67min | 226±41.0sec | 12 |
| LCA$_\%$ | 46.9±1.1sec | 149.3±16.3sec | 10 |
| LSA$_2$ | 12.4±1.1sec | 45.7±4.3sec | 4 |
| DALSA$_2$ | 63.8±14.4sec | 74.4±8.3sec | 6 |

that require tedious full annotations. This alleviates a major obstacle of learning-based methods with regard to clinical applicability and will facilitate the transfer of methods to different clinical domains and settings.

### A. Learning on SUR is Time-efficient

Using SURs saves time during manual creation of the training data. Fewer voxels need to be labeled and the labeling of these voxels is more straightforward, since areas of uncertainty can be avoided. For our experiments these effects add up to an overall reduction of the labeling time by a factor of more than 70.

The use of SURs also reduces the mean training time by a factor greater than 180. The main reason for this – beside the reduction of training data points (voxels) – is the more coherent structure of the data. Since a full segmentation is prone to incorrectly labeled voxels and inconsistent border definitions, the separation of the classes is more difficult. This also explains why lower tree depths perform better if SURs are used.

The usual approach of reducing the training time is to learn only from a randomly drawn subset of all training data. While this approach does not reduce the required labeling time, it does reduce the training time significantly, resulting in times comparable with LSA and DALSA. The times that we reported for DALSA include the calculation time of the correction weights. Since this is an independent step, it could be performed separately from the training. This would reduce the overall training time if multiple training runs were performed – for example during parameter optimization or cross validation.

Learning on sampled training data also reduces the prediction time. Although the effect is not as pronounced as it is on the training time, it still takes twice as long to predict an unseen patient using conventional classifiers compared to using DALSA classifiers. This is mainly due to the more coherent training data, which allow the use of trees with a lower tree depth. The decreased prediction time will be especially important for interactive applications.

### B. Learning on SUR Introduces a Selection Bias

Learning on reduced training data results in a drop in the quality of the prediction results. Neither randomly sampled nor sparsely annotated training data yield classifiers of similar quality as the ones trained on complete annotations (Fig. 6). If the reduction of training data is not done randomly, e.g. when annotating with SURs, a selection bias is introduced. This leads to classifiers which are not optimal for the given problem, as shown in Fig. 8. A simple correction of this effect by an adapted decision threshold is not possible for several reasons: 1. The threshold depends on the unknown $P_{\text{Test}}(x)$ and $P_{\text{Train}}(x)$ and the corresponding $P_{\text{Test}}(y)$ and $P_{\text{Train}}(y)$ and therefore on the rater and his selection of SUR (c.f. Fig. 8). 2. The decision threshold is multi-dimensional for classifiers with more than two classes. 3. Optimal decision thresholds can only be determined for a known gold standard (i.e. fully annotated training data), while the proposed classifier training is bases on SURs only. Fig. 8 could only be plotted because we had complete segmentations available for validation of our approach.

*C. Domain Adaptation Compensates Selection Bias*

The proposed domain adaptation successfully compensates this disproportion of label representations in the training data (Fig. 8). All our experiments show that the use of domain adaptation increases the DICE score and results in a segmentation quality similar to random sampling at comparable ratios. Our experiments with combined SUR sets of different raters show that the level of quality reached by learning from all voxels in the complete annotations can be reached by investing more time into labeling multiple SURs per subject. The DALSA DICE score improves with increasing values of $\lambda$, i.e. with a higher influence of the corrective weights (Fig. 10), further demonstrating the positive effect of DALSA. Our experiment with SVMs and the BraTS challenge data supported these finding.

Training on SURs can increase the classifier's sensitivity for tumorous tissue and at the same time result in an increasing amount of false positive decisions (Fig. 12). This could be caused by the increased tumor-to-tissue ratio in the annotations, as suggested by the finding that domain adaptation lowers the effect when correcting for this ratio. In addition, the classifier's sensitivity could be influenced by the training data quality. Due to ambiguities in the data, complete annotations potentially include a higher number of falsely labeled voxels, resulting in less distinctive label classes. Further betweenclass ambiguities could be caused by healthy tissue voxels that contain inflammation, above-average blood-volume, or chronic stroke and thus have similar appearances to tumorous tissue. The labeling of these voxels would likely be avoided when annotating SURs. The increased amount of false positive decisions that are not connected to the main tumor could likely be reduced by simple post-processing of the results. However, these results could also help identify as yet undiscovered signs of tumorous tissue. The reversed findings in setup III (lower sensitivity and higher positive predictive value for DALSA) are not contradictory, since Kleesiek *et al.* originally sampled

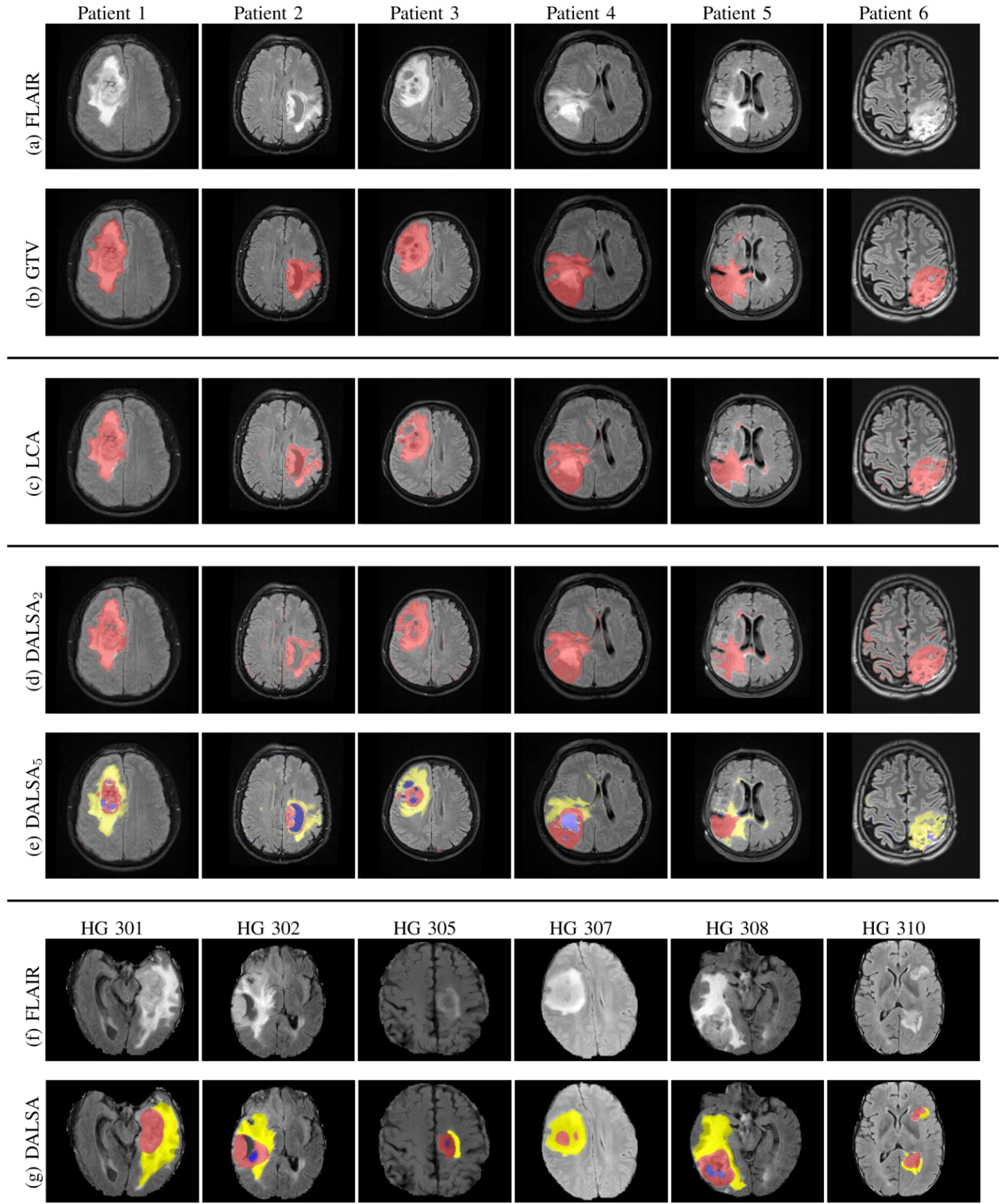

Fig. 7. Exemplary axial slices from setup I (a-e) and setup III (f-g). (a) FLAIR images, (b) gold-standard segmentations, (c) results of classifier trained on complete segmentations, (d) DALSA with 2 classes, and (e) DALSA with 5 classes. In (b-d), the red color indicates 'gross tumor volume'. The color coding in (e) and (g) is: yellow: 'edema', red: 'active tumor', blue: 'necrosis'.

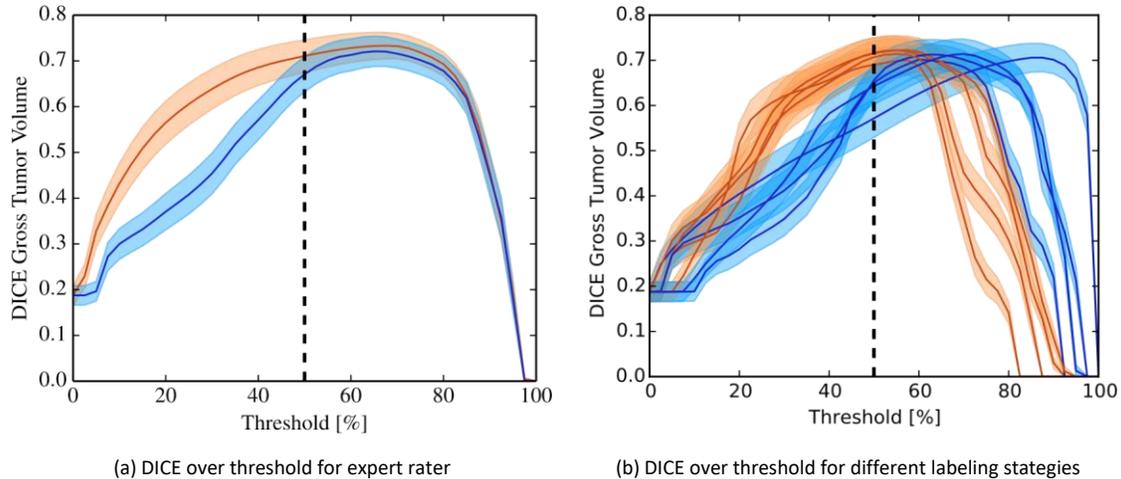

(a) DICE over threshold for expert rater

(b) DICE over threshold for different labeling stategies

Fig. 8. Mean DICE score (line) and standard error (area) for LSA (blue) and DALSA (red) at varying decision thresholds. A balanced curve indicates a well-balanced classifier, while a skewed curve indicates an under- or over-representation of a class. The default threshold is at 50% for a two-class problem. (a): Curves generated from the set of SURs created by the expert rater. (b): Different curves for the different labeling strategies.

the data non-i.i.d., thereby introducing artificial class weights.

The effect of domain adaptation on $LCA_\%$ is only marginal and the absolute improvement is unlikely to be relevant for real applications. The low p-value of the Wilcoxon signedrank test can be explained by the fact that random sampling is never perfectly i.i.d.. Thus our correction does have a minimal effect on each data point, which then adds up to a high sum of ranks in the test.

### D. DALSA Applicable Under Different Conditions

Our experiments demonstrate that DALSA can be easily integrated into existing classification pipelines, such as the one of Kleesiek *et al.*. In cases where a classification algorithm does not offer native support for observation-based weighting, solutions based on classifier ensembles could help and make our approach applicable in combination with virtually any classification algorithm [46]. This is important, since previous studies [47] and our own experiments with SVM demonstrate the impact of the sampling bias also on other classifiers. Our experiments also show a positive effect of DALSA under varying data and feature sets. Using the pipeline of Kleesiek *et al.* [42], we achieved state-of-the-art performance on the basis of sparse training data.

Our experiments also show that the performance of DALSA does depend on the way SURs had been selected (c.f. Fig. 9a). An arbitrarily varying placement of SURs produced the lowest quality end result while at the same time, mostly profiting from the proposed domain adaptation. The advantage of this sampling scheme was its time efficiency: On average it took only 50 seconds to annotate a single patient and add the annotation to the training data. The other annotation schemes (Type 2 – 4) all yielded similar results in terms of DICE scores while requiring an annotation time between two and three minutes per patient. Type 3 has the additional advantage of not requiring tissue annotations close to tissue borders.

Irrespective of the labeling scheme, all SUR-based classifiers could be improved by the use of domain adaptation.

### E. Future Work and Implications

One advantage of our method is that the fast creation of training data allows larger training data sets to be built. We were, for example, able to build a 5-class classifier without much labeling effort. Further tissue classes could be easily added, even without necessarily having to relabel all other tissue classes. It is, for example, possible to add training data for additional stroke or blood classification. The sparse labeling furthermore allows more patients to be labeled within the same time. Based on [7], we think that this will further improve the quality of the trained classifiers and boost the results of our method.

Future work could also be directed to applying crowdsourcing methods in order to further reduce the resources from medical experts when training machine learning algorithms (c.f. [48]). The reduced labeling will reduce the costs caused by the crowd and make this approach more feasible.

## VI. CONCLUSION

We showed that the proposed approach (DALSA) successfully compensates the selection bias introduced by SURs and yields results that are comparable to learning from complete annotations while being significantly more time-efficient. The reduction of labeling time by a factor greater than 70 dramatically eases the establishment of large annotated training collectives with reasonable effort. Our method makes it feasible to create large training sets and update them on a regular basis. It also allows a more efficient adaptation to variations in scanner hardware, modalities and protocols, since labeling of novel training databases takes much less time. We hope that in the future this will support a broader integration and application of automatic tumor segmentation methods in the clinical workflow.

ACKNOWLEDGMENT

We like to thank Sebastian Regnery and Timothy Piotrowski for their valuable help.

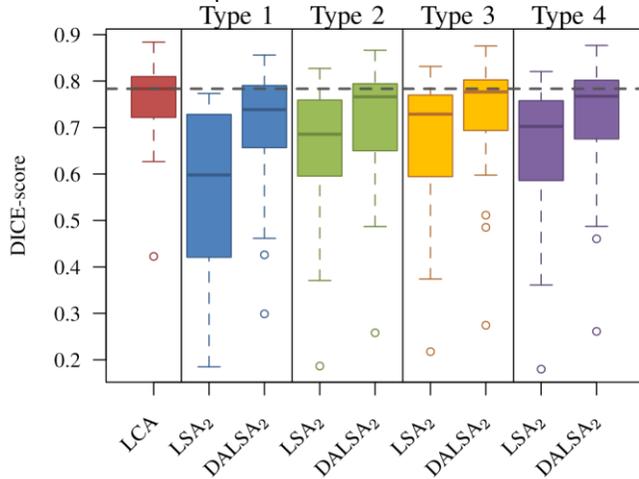
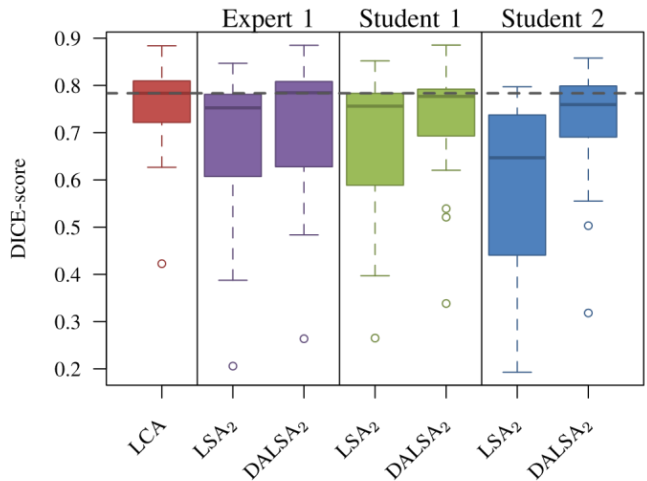
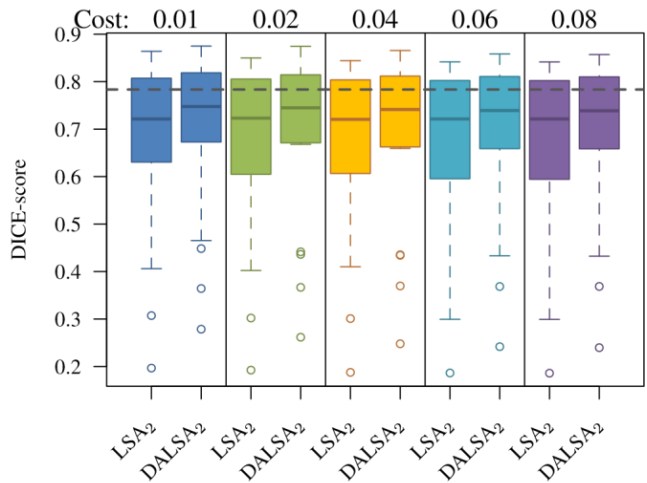

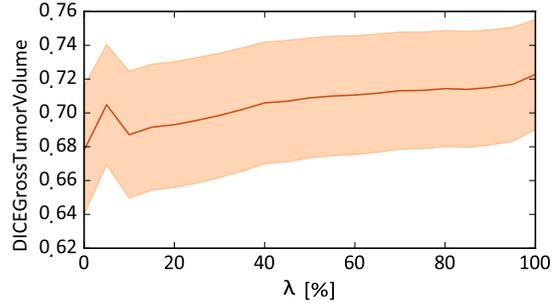

Fig. 9. DICE scores obtained by leave-one-patient-out experiments. (a): Evaluation of different labeling schemes (c.f. Table I). (b): Variability between different raters that drew the SURs independently and blindfolded to the complete tumor segmentation. (c): Comparison of $LSA_2$ and $DALSA_2$ using SVM instead of random forests. The cost factor determines the noise sensitivity.

Fig. 10. Mean and standard error for the leave-one-out experiments acquired with different $\lambda$. Tree depth was set to four in all iterations.

This work was carried out with the support of the German Research Foundation (DFG) as part of project I04 and R01, SFB/TRR 125 Cognition-Guided Surgery.

APPENDIX A

According to Equation 5, the sum over all estimated weights for the SURs of an image can be written as

(9a)

$$\sum_{n_\text{Train}} \hat{w}(x) = \sum_{n_\text{Train}} c \cdot \frac{1-\hat{p}(z=1\mid x)}{\hat{p}(z=1\mid x)} \quad (\hat{p}(z=1\mid x)) \tag{9a}$$

$$= c \cdot n_\text{Train} \cdot \mathbb{E}\left[\frac{1-\hat{p}(z=1\mid x)}{\hat{p}(z=1\mid x)}\right] \tag{9b}$$

$$= c \cdot n_\text{Train} \cdot \frac{1-\mathbb{E}[\hat{p}(z=1\mid x)]}{\mathbb{E}[\hat{p}(z=1\mid x)]} \tag{9c}$$

$$= c \cdot n_\text{Train} \cdot \frac{1 - \frac{n_\text{Train}}{n_\text{Train}+n_\text{Test}}}{\frac{n_\text{Train}}{n_\text{Train}+n_\text{Test}}} \tag{9d}$$

$$= c \cdot \frac{n_\text{Train}}{n_\text{Train}} \cdot (n_\text{Train}+n_\text{Test} - n_\text{Train}) \tag{9e}$$

$$= c \cdot n_\text{Test} \tag{9f}$$

## APPENDIX B

In Fig. 11 we show the ROC curves for the individual subjects. Each curve is obtained by moving the decision threshold for the trained random forest classifier and connecting the individual results.

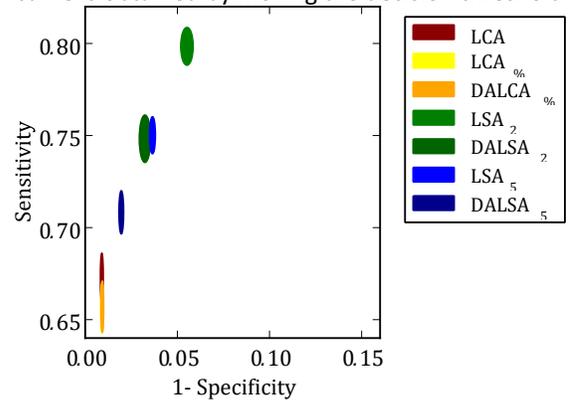

Fig. 12. Specificity and sensitivity of each classifier using the standard decision threshold of 50%. The diameter of the ellipsoids indicates the standard error. LCA$_\%$ is partly covered by DALCA$_\%$.

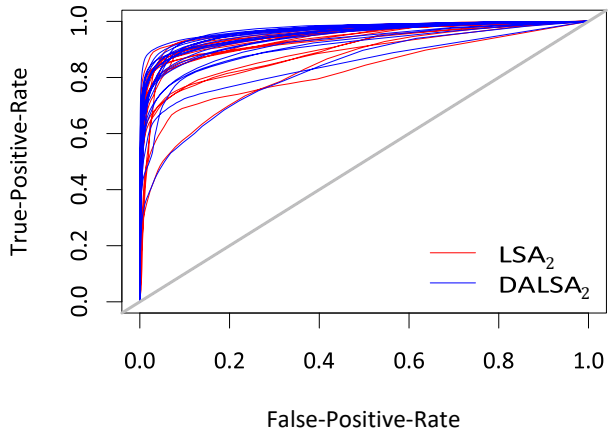

Fig. 11. ROC curves for LSA$_2$ and DALSA$_2$. For each patient, one curve was obtained by training on the other patients and varying the decision threshold of the resulting classifier.